\documentclass[twocolumn]{aastex631}

\newcommand{\abund}[1]{$\log [N({\rm #1})/N({\rm H})]$}

\newcommand{\figref}[1]{Fig.~\ref{#1}}

\newcommand{\ebv}{$E(B-V)$\/}

\newcommand{\kms}{km s$^{-1}$}
\newcommand{\logg}{$\log g$}

\newcommand{\msun}{$M_{\sun}$}

\newcommand{\teff}{$T_{\rm eff}$}

\newcommand{\vrot}{$v \sin i$}

\newcommand{\fuse}{{\em FUSE}}

\shorttitle{Evidence of Third Dredge-Up}
\shortauthors{Dixon}
\graphicspath{{./}{Figures/}}

\begin{document}

\title{Evidence of Third Dredge-Up in Post-AGB Stars in Galactic Globular Clusters}

\correspondingauthor{William V. Dixon}
\email{dixon@stsci.edu}

\author[0000-0001-9184-4716]{William V. Dixon}
\affiliation{Space Telescope Science Institute, 3700 San Martin Drive, Baltimore, MD 21218, USA}



\begin{abstract}

To better understand the mixing and mass loss experienced by low-mass stars as they ascend the asymptotic giant branch (AGB), I have gathered from the literature the abundances of CNO and $s$-process elements in post-AGB stars in Galactic globular clusters.  These species are mixed to the surface during third dredge-up (3DU) events, so their abundance should increase as the star ascends the AGB.  Of the 17 stars in this sample, CNO abundances are available for 11. Of these, four are enhanced in CNO relative to the RGB stars from which they descended, which I take as evidence of 3DU on the AGB.  The enhancement is mainly in the form of carbon.  Of the six stars for which only heavy-element abundances are available, one shows $s$-process enhancements that previous authors have interpreted as evidence of 3DU.  Combining these 17 stars with other recent samples reveals that most globular-cluster post-AGB stars have luminosities $ \log (L/L_{\sun}) \sim 3.25$.  They are the progeny of blue horizontal-branch (HB) stars in clusters with intermediate metallicity ([Fe/H] $\sim -1.5$).  A second group consists of sub-luminous stars associated with high-metallicity clusters ([Fe/H] $\sim -1.0$) with red HBs.  They may be burning helium, rather than hydrogen.  A third group of hot, super-luminous stars is evolving quickly across the Hertzsprung--Russell diagram.  Some of them may be merger remnants.


\end{abstract}



\section{Introduction} \label{sec:intro}

Roughly half of the elements heavier than iron are synthesized in the envelopes of low- and intermediate-mass stars on the asymptotic giant branch (AGB).  In low-mass AGB stars (those with initial mass $M_i$ $<$1.8 \msun), a hydrogen-burning shell provides the bulk of the star's thermonuclear energy.  Periodically, a helium-rich layer below the H-burning shell is ignited, resulting in a thermal pulse (TP), which initiates a series of convection events called third dredge-up (3DU), which in turn brings carbon and $s$-process elements to the stellar surface.  Strong AGB winds then transport these elements into the interstellar medium, where they are incorporated into the next generation of stars.  The chemical evolution of a galaxy thus depends on the activity of its AGB stars.  Unfortunately, the convection efficiencies and mass-loss rates of AGB stars are poorly understood \citep{Herwig:2005}.

The processes of mixing and mass loss on the AGB are the subject of ongoing study.  \citet{Stasinska:2006} compiled a database of 125 post-AGB stars with published photospheric parameters (effective temperature and surface gravity) and chemical compositions to search for trends with mass and metallicity.  More recently, \citet{Ventura:2017} examined the chemical abundances of 142 Galactic planetary nebulae (PNe) to test models of AGB evolution and dust formation, and \citet{Marigo:2022} used the initial--final mass relation of white dwarfs to constrain the efficiency of 3DU in solar-metallicity AGB stars.

Another approach is to study the chemical abundances of UV-bright stars in globular clusters.  Stars brighter than the horizontal branch (HB) and bluer than the red giant branch (RGB) represent a variety of post-HB evolutionary paths.  Red HB (RHB) stars with sufficiently massive envelopes fully ascend the AGB before becoming true post-AGB stars.  Blue HB (BHB) stars with less-massive envelopes partially ascend the AGB before peeling off to become post-Early AGB (PEAGB) stars.  The bluest HB stars (extreme HB or EHB stars), with the least-massive envelopes, do not attempt an AGB ascent, instead evolving directly toward the tip of the white-dwarf cooling curve.  Stars in globular clusters offer the additional benefits of known distance, reddening, and initial chemical composition.

As a star ascends the AGB, its envelope is depleted from above by the stellar wind and from below by the H-burning shell.  When the mass of the envelope is no longer sufficient to support nuclear burning, the star leaves the AGB, moves across the HR diagram at constant luminosity, and eventually settles on the white-dwarf cooling curve.  If the star departs the AGB before reaching its tip, then one would expect its abundances to reflect the extent of its enrichment due to 3DU and its luminosity to record exactly how far up the AGB the star ascended.  \citet{Gonzalez:Wallerstein:1994} searched for this effect by examining five UV-bright stars in the globular cluster $\omega$ Cen.  Using high-resolution ground-based spectra, they derived the stars' effective temperatures, surface gravities, and the abundances of 26 elements.  They found that the most luminous stars display significant enhancements of the CNO and $s$-process elements, with the CNO abundance rising abruptly in stars with luminosity $\log (L/L_{\sun}) > 2.7$.  

In this paper, I combine the approach of \citet{Gonzalez:Wallerstein:1994} with that of \citet{Stasinska:2006}, compiling a list of post-AGB stars in globular clusters with published photospheric parameters and abundances and searching for trends in the frequency of 3DU as a function of stellar mass and luminosity.  In Section \ref{sec_theory}, I review the theory of post-main sequence evolution and the expectations for mixing and mass loss at each evolutionary stage.  In Section \ref{sec_data}, I present my sample of stars and their parameters and abundances.  In Section \ref{sec_analysis}, I search for evidence of 3DU and discuss the resulting trends.  In Section \ref{sec_discussion}, I discuss several individual stars and consider other recent results.  I present my conclusions in Section \ref{sec_conclusions}.

\section{Theory}\label{sec_theory}

A star's atmospheric abundances are modified by its post-main sequence evolution.  According to stellar-evolution theory \citep{Iben:Renzini:83}, the arrival of a low-mass star on the red giant branch is accompanied by a deepening of its convective envelope, which brings to the surface the ashes of hydrogen burning via the CNO cycle.  This process, known as first dredge-up, results in a doubling of the surface $^{14}$N abundance, a reduction in the $^{12}$C abundance of about 30\%, and practically no change in the abundance of $^{16}$O.

No further mixing on the RGB is predicted by theory, but observations of low-mass red giants in both clusters and the field \citep[e.g.,][]{Gratton:2000} suggest that some form of mixing continues to bring CNO-processed material to the surface.  The carbon abundance continues to fall, and the nitrogen abundance to rise, as a star ascends the RGB \citep{Smith:2002}.  While the CNO process converts carbon into nitrogen, the star's total CNO abundance is unchanged.

No mixing is expected to occur on the horizontal branch.  A second episode of dredge-up can occur at the base of the AGB, but only in stars more massive than those present in globular clusters today.  Third dredge-up occurs (in some stars) as they ascend the AGB and experience thermal pulses in their helium-burning shells.  Third dredge-up brings $^{12}$C (produced via the triple-$\alpha$ process), a small amount of $^{16}$O (produced via the reaction $^{12}$C($\alpha, \gamma)^{16}$O), and $s$-process isotopes into the convective envelope.  The envelope becomes more carbon-rich, and the total CNO abundance rises.  The efficiency of 3DU increases with core and envelope mass and with decreasing metallicity \citep{Herwig:2005}.

When looking for evidence of 3DU, an important consideration is the mass of the stellar envelope, which is reduced from above by the stellar wind and from below by the advancing H-burning shell as the star ascends the AGB.  Stellar-evolution models predict that, near the end of the AGB, the envelope mass can change by more than an order of magnitude during a single inter-pulse period \citep{Vassiliadis:Wood:1993, Vassiliadis:Wood:1994}, so a given amount of enriched material has a much greater impact on the photospheric abundances at the end of star's AGB journey than at the beginning.  The detection of 3DU material thus constrains the efficiency of the mixing and mass-loss processes at work on the AGB.

\section{Data}\label{sec_data}

Several compendia of UV-bright stars in globular clusters have been published over the years.  \citet[hereafter ZNG]{ZNG:1972} presented finding charts for UV-bright stars in 27 clusters, and \citet{Harris:1983} compiled a list of probable members in 29 clusters.  \citet{Moehler:2019} determined the photospheric parameters for seven post-AGB stars and collected the parameters of 17 more from the literature.  \citet{Bond:2021} extracted the coordinates, parallaxes, and proper motions of the entire ZNG sample from the Gaia Early Data Release 3 \citep[EDR3;][]{Gaia_Mission, GaiaEDR3} and determined which objects are probable cluster members.  Finally, \citet{Davis:2022} conducted a survey of post-horizontal-branch stars in 97 Galactic globular clusters and identified 13 stars as post-AGB stars; \citet{Ciardullo:2022} refined this sample to 10 stars with $M_V < -3.0$.

\begin{deluxetable*}{lclrcccDccl}
\tablecaption{Post-AGB Stars in Globular Clusters \label{tab:stars}}
\tablehead{
\colhead{Cluster} & \colhead{Alt.} & \colhead{Star} & \colhead{\teff} & \colhead{\logg} & \colhead{[Fe/H]} & \colhead{$V$} & \multicolumn2c{$\log (L/L_{\sun})$} & \colhead{$M/M_{\sun}$} & \colhead{3DU?} & \colhead{Refs.} \\
& \colhead{Name} & & \colhead{(K)} & & & \colhead{(mag)} 
}
\decimals
\startdata
NGC 104 &       47 Tuc   &       Bright Star     &       10,850  &       2.20    &    $-0.94$   &   10.637  &       3.06\tablenotemark{a}    &       0.53    &       No      &	1, 2 \\
NGC 1904        &       M79     &       PAGB\tablenotemark{b}    &       6300    &       0.80    &     $-2.08$   &     12.215  &       3.28    &       0.31    &       No\tablenotemark{c}      &	3, 2 \\
NGC 5139        &       $\omega$ Cen    &       ROA 24, HD 116745       &       6250    &       1.00    &     $-1.97$   &   10.785  &       3.22    &       0.44    &       Yes     &	4, 2 \\
        &           &       ROA 5701        &       25,000   &       3.30    &     $-2.71$     &   13.071  &       3.29    &       0.41    &       No      &	5, 2 \\
        &           &       V1, LEID 32029  &       5610    &       0.89    &    $-1.69$   &       10.829  &       3.24    &       0.55    &       Yes     &	4, 6 \\
        	&	     &     V29, LEID 43105 &	5353	&	0.81	&	$-1.91$     &    12.015	&	2.79	&	0.20	&	Yes	&	4, 6 \\
NGC 5272        &       M3      &       ZNG 1, vZ 1128  &       36,600   &       3.95     &      $-1.79$    &       14.991  &       3.32    &       0.42    &       No      &	7, 2 \\
NGC 5904        &       M5      &       V42	&	5454	& 	0.84	&	$-1.22$    &    11.19\phn	&	3.28	&	0.60	&	No\tablenotemark{c}	&   	8, 9 \\
        &               &       V84     &       5310    &       0.97    &      $-1.22$      &       11.42\phn  &       3.20    &       0.75    &       No\tablenotemark{c}      &	10, 9 \\
NGC 5986        &      \nodata        &       PAGB 1, ID 6    &       8750    &       2.00    &    $-1.74$     &       12.569  &       3.29    &       1.34    &       No\tablenotemark{c}      &	11, 2 \\
       &               &       PAGB 2, ID 7    &       6300    &       1.00    &    $-1.84$    &       12.664  &       3.25    &       0.45    &       Yes\tablenotemark{c}    &	11, 2 \\
NGC 6205        &       M13     &       ZNG 1, Barnard 29       &       21,400   &       3.10    &     $-1.55$     &       13.093  &       3.29\tablenotemark{a}    &       0.48    &       No      &	12, 2 \\
NGC 6254        &       M10     &       ZNG 1, I-33     &       27,000   &       3.60    &      $-1.12$    &       13.528  &       3.33    &       0.65    &       No      &	13, 2 \\
NGC 6397        &       \nodata        &       ROB 162 &       51,000   &       4.50    &     $-1.78$     &      13.186  &       3.44    &       0.52    &       No      &	14, 15 \\
NGC 6712        &       \nodata        &       ZNG 1, C26   &       11,000   &       2.10    &      $-1.54$    &       13.211  &       3.03    &       0.37    &       No\tablenotemark{c}     &		11, 13, 2 \\
NGC 7078        &       M15     &       ZNG 1, K559     &       28,000   &       3.70    &    \nodata    &       15.043  &       3.19    &       0.51    &       No      &	13, 2 \\
        &            &       K648    &       36,360   &       3.96    &     $-2.27$    &      14.320  &       3.49\tablenotemark{a}    &       0.66    &       Yes     &	16, 2 \\
\enddata
\tablenotetext{a}{Luminosity from first reference, rescaled to the cluster distance listed in Table \ref{tab:clusters}.}
\tablenotetext{b}{Also known as 2MASS J05241036-2429206.  See \citet{Bond:2016}.}
\tablenotetext{c}{CNO abundance not available.  3DU status is based on heavy-element abundances and is taken from the first reference (first and second references for ZNG 1 in NGC 6712).}
\tablecomments{For the stars that are not variable, the effective temperature and surface gravity are taken from the first reference and the $V$ magnitude from the second.   Stellar parameters of variable stars are estimated from intensity-weighted photometry as discussed in the text.  [Fe/H] is from the first reference, except for ROA 5701 in $\omega$ Cen, which is from \citet{Moehler:98}.  [Fe/H] is adjusted to the solar value of \citet{Asplund:2009}.}
\tablerefs{(1) \citet{Dixon:2021}, (2) \citet{Stetson:2019}, (3) \citet{Sahin:2009}, (4) \citet{Gonzalez:Wallerstein:1994}, (5) \citet{Thompson:2007}, (6) \citet{Braga:2020}, (7) \citet{Chayer:2015}, (8) \citet{Carney:1998}, (9) \citet{Rabidoux:2010}, (10) \citet{Gonzalez:Lambert:1997}, (11) \citet{Jasniewicz:2004}, (12) \citet{Dixon:2019},  (13) \citet{Mooney:2004}, (14) \citet{Heber:Kudritzki:86}, (15) \citet{Zacharias:2013}, (16) \citet{Otsuka:2015}}
\end{deluxetable*}

\citet{Moehler:2019} divided the UV-bright stars into three groups.  The most luminous, with $\log (L/L_{\sun}) \ga 3.1$, fully ascended the AGB.  The second group, with $2.65 \la \log (L/L_{\sun}) \la 3.1$, consists of post-EAGB and post-EHB stars.  The third group, with $1.8 \la \log (L/L_{\sun}) \la 2.65$, is mostly post-EHB stars, with a handful of stars evolving from the hot end of the BHB towards the AGB.  To identify all stars that experienced even a few thermal pulses, I selected stars from the first two groups, that is, all UV-bright stars with luminosity $\log (L/L_{\sun}) > 2.65$.

\begin{deluxetable*}{lcccDD}
\tablecaption{Properties of Globular Clusters \label{tab:clusters}}
\tablehead{
\colhead{Cluster} & \colhead{Alt. Name} & \colhead{[Fe/H]} & \colhead{\ebv} & \multicolumn2c{Distance (kpc)} & \multicolumn2c{HBR}
}
\decimals
\startdata
NGC 104	&	 47 Tuc	&	 $-0.72$	&	0.04	&	4.521	&	 $-0.99$ \\ 
NGC 1904	&	M79	&	$-1.60$	&	0.01	&	13.078	&	0.89 \\
NGC 5139	&	 $\omega$ Cen	&	 $-1.53$	&	0.12	&	5.426	&	 0.89 \\ 
NGC 5272	&	 M3	&	 $-1.50$	&	0.01	&	10.175	&	 0.08 \\ 
NGC 5904	&	 M5	&	 $-1.29$	&	0.03	&	7.479	&	 0.37 \\ 
NGC 5986	&	\nodata	&	$-1.59$	&	0.28	&	10.540	&	0.95 \\
NGC 6205	&	 M13	&	 $-1.53$	&	0.02	&	7.419	&	 0.97 \\ 
NGC 6254	&	 M10	&	 $-1.56$	&	0.28	&	5.067	&	 0.94 \\ 
NGC 6397	&	 \nodata	&	 $-2.02$	&	0.18	&	2.482	&	 0.93 \\ 
NGC 6712	&	\nodata	&	$-1.02$	&	0.45	&	7.382	&	$-0.60$ \\
NGC 7078	&	 M15	&	 $-2.37$	&	0.10	&	10.709	&	 0.72 \\ 
\enddata
\tablecomments{[Fe/H] and \ebv\ from \citet{Harris:96, Harris:2010}. Distance from \citet{Baumgardt:2021}.  Horizontal-branch ratio (HBR) from \citet{Borkova:Marsakov:2000} for 47~Tuc and $\omega$ Cen, \citet{Lee:1994} for the rest.}
\end{deluxetable*}

Using the above sources, I identified some 30 post-AGB stars in globular clusters and searched the literature for their photospheric parameters and chemical abundances.  I found CNO abundances (or upper limits) for 11 stars and heavy-element abundances for six more.  The complete sample of 17 stars is presented in Table \ref{tab:stars}.  Various parameters of their parent clusters are presented in Table \ref{tab:clusters}.

\citet{Stetson:2019} provide ground-based Johnson-Cousins UBVRI photometry for 48 Galactic globular clusters, with modern calibrations and typical uncertainties of a few millimagnitudes.  I use these values for all but five of the stars.  For the variable stars in NGC 5139 ($\omega$ Cen) and NGC 5904 (M5), I use the intensity-weighted mean magnitudes from \citet{Braga:2020} and \citet{Rabidoux:2010}, respectively.  ROB~162 in NGC~6397 is not included in the \citeauthor{Stetson:2019} catalog, so I use a measurement from \citet{Zacharias:2013}.

For the stars in Table \ref{tab:stars} that are not variable, the effective temperature and surface gravity are taken from the first reference in the list and the $V$ magnitude from the second.  If the authors computed the stellar luminosity by scaling a synthetic spectrum to their data, then I rescaled their value using the cluster distance from \citet{Baumgardt:2021}.  If not, I derived the luminosity from the $V$ magnitude as follows.  Taking the star's effective temperature and surface gravity from Table \ref{tab:stars} and the cluster metallicity from Table \ref{tab:clusters}, I computed the bolometric correction ($BC_V$) using the MASH3 program provided by \citet{Worthey:2011}.  I then used the foreground reddening and cluster distance from Table \ref{tab:clusters} to convert the star's $V$ magnitude ($m_V$) to an absolute magnitude ($M_V$) and thence to a luminosity using the standard equations:
\begin{equation}
M_V = m_V - (5 \times \log D - 5 + A_V)
\end{equation}
and
\begin{equation}
\log \frac{L}{L_{\sun}} = -0.4 \times (M_V + BC_V - 4.74),
\end{equation}
where $M_V$, $m_V$, and $BC_V$ are in magnitudes, the distance $D$ is in parsecs, and $A_V = 3.1 \times E(B-V)$.  The stellar mass was computed from the relation
\begin{equation}
\log \frac{M}{M_{\sun}} = \log \frac{L}{L_{\sun}} + \log \frac{g}{g_{\sun}}  - 4 \times \log \frac{T}{T_{\sun}} .
\end{equation}

\begin{deluxetable*}{lclcccccc}
\tablecaption{CNO Abundances of Sample Stars \label{tab:cno}}
\tablehead{
\colhead{Cluster} & \colhead{Alt. Name} & \colhead{Star} & \colhead{$\epsilon$(C)} & \colhead{$\epsilon$(N)} & \colhead{$\epsilon$(O)} & \colhead{$\epsilon$(C+N+O)$_*$} & \colhead{$\epsilon$(C+N+O)$_{cl}$} & \colhead{Refs.}
}
\startdata
NGC 104 & 47 Tuc & Bright Star & $ 6.74 \pm 0.13 $ & $ 8.40 \pm 0.31 $ & $ 8.24 \pm 0.15 $ & $ 8.63 \pm 0.19 $ & $ 8.75 \pm 0.13 $ & 1, 2 \\ 
NGC 5139 & $\omega$ Cen & ROA 24 & $ 7.13 \pm 0.05 $ & $ 7.77 \pm 0.06 $ & $ 8.20 \pm 0.04 $ & $ 8.36 \pm 0.03 $ & $ 7.43 \pm 0.10 $ & 3, 4 \\ 
  & & ROA 5701 & $ < 6.54 $ & $ 7.06 \pm 0.12 $ & $ 7.76 \pm 0.10 $ & $ 7.84 \pm 0.09 $ & $ 7.69 \pm 0.19 $ & 5, 4 \\ 
  & & V1 & $ 7.47 \pm 0.11 $ & $ (7.4) $ & $ 8.32 \pm 0.11 $ & (8.4) & $ 7.69 \pm 0.19 $ & 3, 4 \\ 
  & & V29	& $ 6.77 \pm	0.21	$ & $	(7.4)	$ & $	8.07	\pm	0.24	$ & (8.2) & $ 7.43 \pm 0.10 $ &	3, 4 \\					
NGC 5272 & M3 & vZ 1128 & $ 6.41 \pm 0.13 $ & $ 7.56 \pm 0.15 $ & $ 7.47 \pm 0.19 $ & $ 7.83 \pm 0.11 $ & $ 7.87 \pm 0.16 $ & 6, 2 \\ 
NGC 6205 & M13 & Barnard 29 & $ 6.03 \pm 0.09 $ & $ 7.49 \pm 0.13 $ & $ 7.54 \pm 0.11 $ & $ 7.82 \pm 0.08 $ & $ 7.92 \pm 0.08 $ & 7, 2 \\ 
NGC 6254 & M10 & ZNG 1 & $ 5.95 \pm 0.20 $ & $ 7.34 \pm 0.15 $ & $ 7.66 \pm 0.15 $ & $ 7.84 \pm 0.11 $ & $ 8.02 \pm 0.19 $ & 8, 2 \\ 
NGC 6397 & \nodata & ROB 162 & $ < 5.50 $ & $ 6.96 \pm 0.45 $ & $ 6.85 \pm 0.41 $ & $ 7.21 \pm 0.31 $ & $ 7.40 \pm 0.04 $ & 9, 10, 11 \\ 
NGC 7078 & M15 & K559 & $ < 6.83 $ & $ 6.91 \pm 0.08 $ & $ 6.32 \pm 0.20 $ & $ 7.01^{+0.22}_{-0.08} $ & $ 6.96 \pm 0.04 $ & 8, 12 \\ 
 & & K648 & $ 9.38 \pm 0.02 $ & $ 6.53 \pm 0.10 $ & $ 8.36 \pm 0.10 $ & $ 9.42 \pm 0.02 $ & $ 6.96 \pm 0.04 $ & 13, 12 \\
\enddata
\tablerefs{(1) \citet{Dixon:2021}, (2) \citet{Meszaros:2020}, (3) \citet{Gonzalez:Wallerstein:1994}, (4) \citet{Marino:2012}, (5) \citet{Thompson:2007}, (6) \citet{Chayer:2015}, (7) \citet{Dixon:2019}, (8) \citet{Mooney:2004}, (9) \citet{Chayer:Dixon:2015}, (10) \citet{Carretta:2005}, (11) \citet{Lind:2011}, (12) \citet{Sneden:1997}, (13) \citet{Otsuka:2015}}
\end{deluxetable*}

For the variable stars, whose photospheric parameters vary with luminosity, I use the magnitude-weighted $B - V$ colors from the references given in Table \ref{tab:stars} to estimate the stars' effective temperature, surface gravity, and bolometric correction.  I assume that the stars are evolving along the post-AGB evolutionary track \citep[taken from][]{Moehler:2019} for a star with [Fe/H] = $-1.5$ and zero-age HB (ZAHB) mass 0.55 \msun\ (chosen because it is the lowest-mass track that ascends the AGB).  Each point on the track represents a locus on the \teff\ vs.\ \logg\ plane.  At each point, I use the MASH3 program \citep{Worthey:2011} to estimate the corresponding values of $(B - V)_0$ and $BC_V$.  Given the stellar value of $(B - V)_0$, I interpolate along the track to find the corresponding values of \teff, \logg, and $BC_V$.

The column labeled ``3DU?'' in Table \ref{tab:stars} indicates whether or not the star shows evidence of third dredge-up.  For six stars, heavy-element abundances are available, but CNO abundances are not. For these stars, the value in this column is ``Yes'' for stars whose photospheres are enhanced in $s$-process species.  For the other stars, the values are derived from their CNO abundances in Section \ref{sec_analysis}.

For the 11 stars for which CNO abundances are available, Table \ref{tab:cno} presents the carbon, nitrogen, and oxygen abundances of each star and its parent cluster.\footnote{This paper employs the standard nomenclature: the abundance of element X is given by $\epsilon$(X) = \abund{X} + 12, and [X/Y] = $\log [N({\rm X})/N({\rm Y})]_{\star} - \log [N({\rm X})/N({\rm Y})]_{\sun}$.}  The stellar values are taken from the first reference in the list and the cluster values from the second (and third, if present). 
For three stars, only upper limits on the carbon abundance are available.  For ROA~5701 in NGC~5139 ($\omega$ Cen) and ROB~162 in NGC~6397, the difference between the upper limit to $\epsilon$(C+N+O) and the measured value of $\epsilon$(N+O) is less than the uncertainty in the latter, so I list $\epsilon$(N+O) in Table \ref{tab:cno}.  For K559 in NGC~7078 (M15), $\epsilon$(N+O) = $7.01 \pm 0.08$, while $\epsilon$(C+N+O) $< 7.23$, so I quote $\epsilon$(C+N+O) $= 7.01^{+0.22}_{-0.08}$.  Abundances are often quoted relative to the solar value.  I present all abundances as ratios relative to hydrogen.  Most authors provided references to their adopted solar values.  When they did not, I used references from their previous papers or close collaborators.

Globular clusters consist of multiple stellar populations with a range of light-element abundances \citep{Bastian:Lardo:2018}.  Even so, in most clusters the combined abundance of carbon, nitrogen, and oxygen, $\epsilon$(C+N+O), is roughly constant \citep{Smith:1996, Smith:1997, Ivans:1999, Carretta:2005}.  The mean value of $\epsilon$(C+N+O) thus is a reasonable estimate of the star's CNO abundance while on the RGB.

\citet{Meszaros:2020} derive the abundance of 11 elements in 2283 red-giant stars in 31 globular clusters from high-resolution spectra obtained by the SDSS-IV APOGEE-2 survey.  Using their catalog, I compute the mean value of $\epsilon$(C+N+O) for each of 47 Tuc, M3, M10, and M13.  For $\omega$ Cen, I employ the results of \citet{Marino:2012}, who found a tight correlation between $\epsilon$(C+N+O) and [Fe/H].  The stars ROA 24 and V29 fall in the group with [Fe/H] $< -1.9$, for which the mean value of $\epsilon$(C+N+O) is $7.43 \pm 0.10$.  The star V1 falls in the group with $-1.9 <$ [Fe/H] $< -1.65$, for which $\epsilon$(C+N+O) is $7.69 \pm 0.19$.  The iron abundance of ROA~5701 is even lower than that of ROA~24, but I include it in the V1 group for the reasons discussed in Section \ref{sec_omegacen}.  For NGC~6397, I combine the mean abundance of C and N \citep{Carretta:2005} with that of O \citep{Lind:2011} for the cluster's second-generation stars.  Finally, for M15, I adopt the mean $\epsilon$(C+N+O) value for five giant stars studied by \citet{Sneden:1997}.  For all but NGC~6397, the error bars represent the scatter in the stellar $\epsilon$(C+N+O) values, not the error in the mean.

\begin{figure}
\plotone{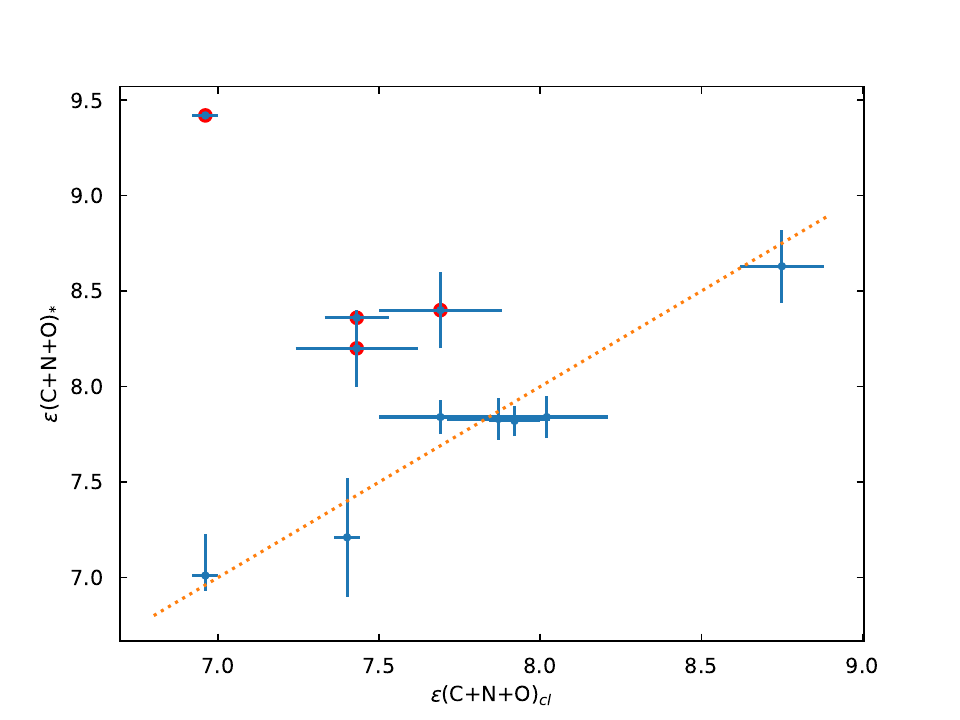}
\caption{CNO abundance of star versus mean CNO abundance of host cluster.  Stars exhibiting evidence of 3DU are highlighted in red.  The dotted line indicates a ratio of unity.}
\label{fig_cno}
\end{figure}

\section{Analysis}\label{sec_analysis}

Consider the 11 stars with CNO abundances listed in Table \ref{tab:cno}.  In their survey of post-AGB stars, \citet{Stasinska:2006} defined objects that have experienced 3DU as those in which the ratio (C+N+O)/S is larger than in the Sun.  They divided by the sulfur abundance to remove the effects of abundance gradients across the Galaxy.  I define 3DU stars as objects for which $\epsilon$(C+N+O) for the star is greater than $\epsilon$(C+N+O) for the RGB stars from which it descended.  

\begin{figure*}
\epsscale{1.1}
\plotone{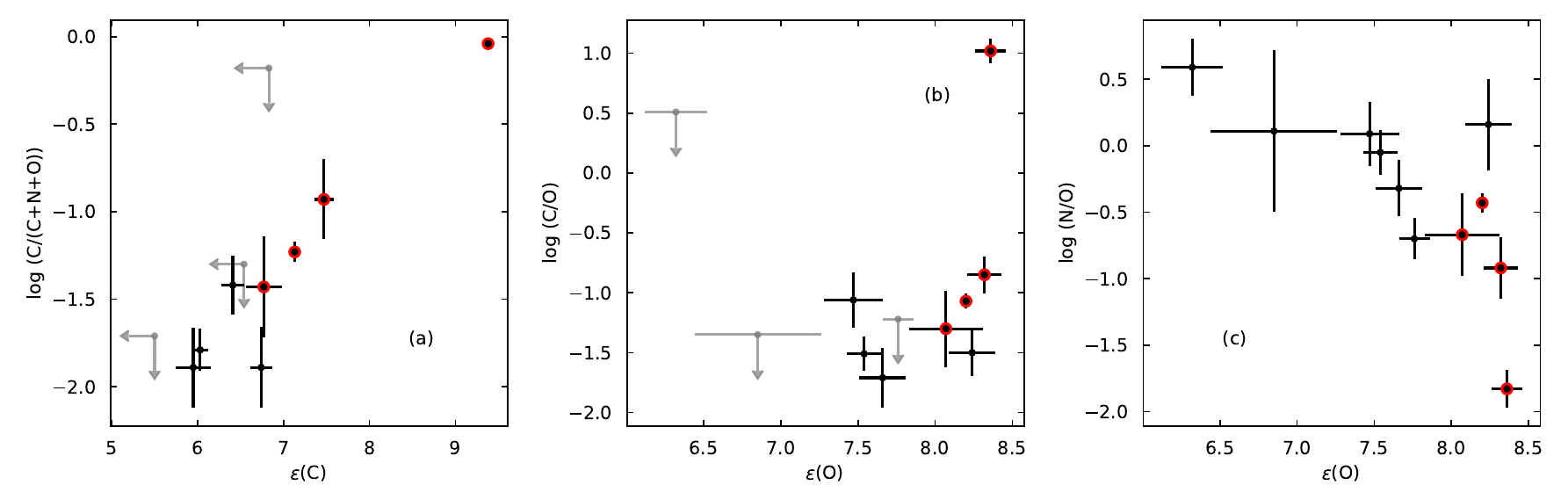}
\caption{Abundance ratios of the stars with measured CNO abundances.  Upper limits are plotted in grey.  Stars exhibiting evidence of 3DU are highlighted in red.  (a) Fraction of CNO in the form of carbon versus carbon abundance.  (b) Carbon-to-oxygen ratio versus oxygen abundance.  (c) Nitrogen-to-oxygen ratio versus oxygen abundance.}
\label{fig_ratios}
\end{figure*}

In \figref{fig_cno}, the stellar value of $\epsilon$(C+N+O) is plotted against the mean value of its parent cluster.  The dotted line indicates a ratio of unity.  Four stars, ROA 24, V1, and V28 in $\omega$ Cen and K648 in M15, are enhanced in CNO; they are highlighted in red and indicated in Table \ref{tab:stars}.  Two stars, ROA~5701 in $\omega$ Cen and K559 in M15, lie just above the dotted line, but given the scatter, I do not consider them to be enhanced.  The enhancement appears to be mostly in the form of carbon.  In panel (a) of \figref{fig_ratios}, C/(C+N+O), the fraction of CNO in the form of carbon, is plotted against the stellar carbon abundance.  As the carbon abundance rises, the fraction of CNO in the form of carbon approaches unity.  As pointed out by \citet{Stasinska:2006}, this trend suggests a scenario in which primary carbon, produced by the triple-$\alpha$ reaction, is brought to the stellar surface during 3DU.  In panel (b), the ratio C/O is plotted against the oxygen abundance of each star.  Note that, by my definition, an object need not be a carbon star (one in which the photospheric ratio of C/O $> 1$) to exhibit evidence of 3DU.  Only in K648, the central star of the planetary nebula in M15, does the abundance of carbon exceed that of oxygen.  
 
In panel (c) of \figref{fig_ratios}, N/O is plotted against the oxygen abundance.  An anti-correlation between these quantities has been seen among some post-AGB stars in the Galaxy \citep{Stasinska:2006} and planetary nebulae in the Magellanic Clouds \citep{Henry:1989}.  It is generally attributed to the production of N at the expense of O (via the ON cycle) in material that is later brought to the stellar surface by the second dredge-up.  This explanation is not appropriate for stars in globular clusters, as they do not undergo second dredge-up.  The O abundance is highest---and the N/O ratio lowest---among the four 3DU stars, suggesting that 3DU has brought both C and O to the surface.

\begin{figure*}
\epsscale{0.75}
\plotone{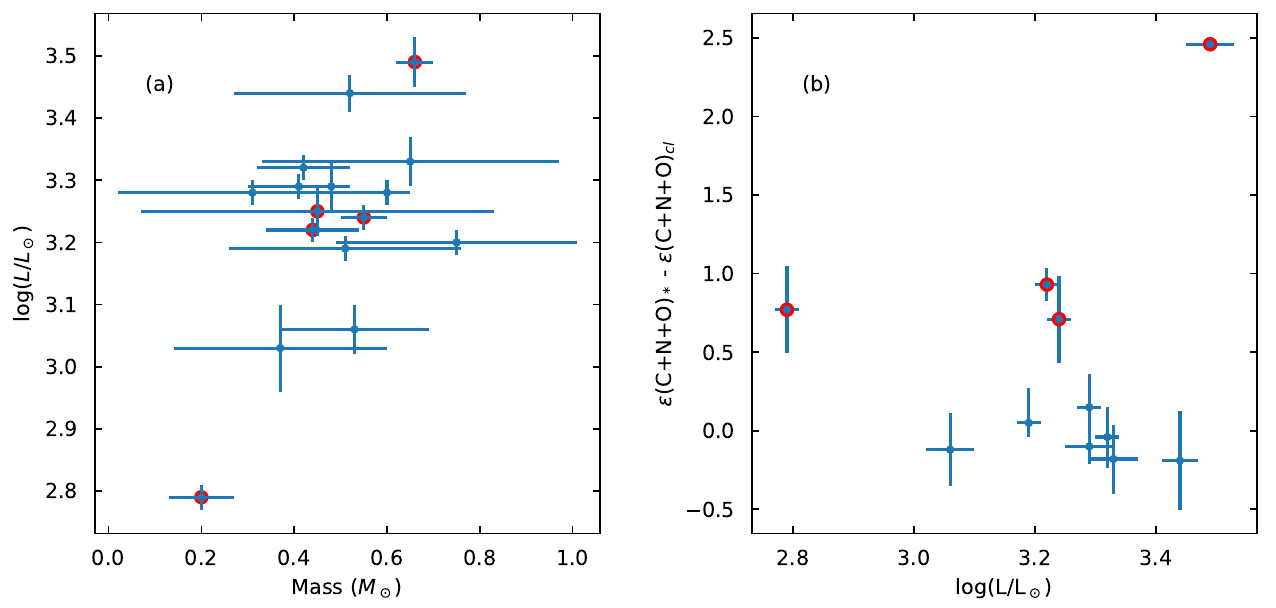}
\caption{(a) Luminosity versus mass for 16 of the sample stars.  (b) Stellar minus cluster value of $\epsilon$(C+N+O) versus luminosity.  Only the stars with CNO abundances are shown.  In both panels, stars exhibiting evidence of 3DU are highlighted in red.}
\label{fig_mass}
\end{figure*}

Stellar-evolution theory suggests that, for post-AGB stars, stellar luminosity is correlated with mass, and evidence of 3DU is correlated with both parameters, the idea being that only the most massive stars have envelopes sufficient to power their full ascent of the AGB, and these stars leave the AGB with greater luminosities than lower-mass objects.  This hypothesis is explored in \figref{fig_mass}.  In panel (a), luminosity is plotted against stellar mass for 16 of the 17 stars listed in Table \ref{tab:stars}.  The star PAGB~1 in NGC~5986 is omitted, as its mass (formally, $1.34 \pm 1.26$ \msun) is poorly constrained.  In panel (b) the CNO enhancement $\epsilon$(C+N+O)$_* - \epsilon$(C+N+O)$_{cl}$ is plotted as a function of luminosity for the 11 stars with CNO abundances.  In both panels, the error bars represent the measurement uncertainties propagated through equations (1) through (3).  

The stellar masses plotted in panel (a) show considerable scatter, but the error bars are large.  Most stars have masses consistent with 0.53 \msun, the mean mass of white dwarfs forming in globular clusters today \citep{Kalirai:2009}.  Two notable exceptions, V29 in $\omega$ Cen (the least luminous star) and K648 in M 15 (the most luminous), are discussed below.  Four of the five 3DU stars have relatively small error bars; they exhibit a strong correlation between luminosity and mass.  In panel (b), three of the 3DU stars show CNO enhancements of $0.5 - 1.0$ dex, while K648 is enhanced by nearly 2.5 dex; whether K648 is unique or indicates a break in the CNO-luminosity relation cannot be determined from the present data.

\section{Discussion}\label{sec_discussion}

The stars that show evidence of 3DU have been studied in some detail.  In this section, I review the literature on each of them, discuss a few of the non-3DU objects, and compare my stars with two recently-published samples.

\subsection{The UV-Bright Stars in NGC 5139 ($\omega$ Cen)}\label{sec_omegacen}

\citet{Gonzalez:Wallerstein:1994} obtained high-resolution spectra of five UV-bright stars in $\omega$ Cen.  In order of decreasing luminosity, the stars are ROA 24, V1, V29, ROA 342, and V48.  The authors derived the stars' effective temperatures, surface gravities, and the abundances of 26 elements.  Their goal was to determine the AGB luminosity at which the products of 3DU reach the stellar surface.  They found that the three most luminous stars are significantly enhanced in CNO, corresponding to a major transport of helium-burning products between $\log (L/L_{\sun}) = 2.51$ and 2.74 (using their luminosity estimates for ROA 342 and V29).  An excess of $s$-process elements was seen in these three stars, along with V48, but not in ROA 243.  The authors suggested that V48 is in a transition stage in which $s$-process elements have reached the surface before an appreciable enhancement of He-burning products.    
 
The low mass and luminosity of V29 suggest that it is not an ordinary post-AGB star.  \citet{Gonzalez:Wallerstein:1994} postulated that both V1 and V29 are ``blue loop'' stars that have temporarily moved from the AGB into the instability strip of the Population II Cepheids in response to a thermal pulse.  Such objects may return to---and continue to climb---the AGB, so V29 may not have achieved its final luminosity.  Alternatively, the star's low mass suggests that it may be the product of an interaction with a binary companion.  By overflowing its Roche lobe during its ascent of the AGB, the star could have lost most of its mass, perhaps revealing the products of helium burning in the process; however, the fact that the star's abundance patterns match those of the other 3DU stars in \figref{fig_ratios} suggests that its evolution was not terribly different from theirs.

The star ROA 5701 has an unusual iron abundance.  \citet{Moehler:98} derive [Fe/H] = $-2.72 \pm 0.11$, making it more metal-poor than any star in the catalog of \citet{Marino:2012} by some 0.7 dex.  The authors suggest that the iron depletion could be explained by gas-dust separation in the AGB progenitor's atmosphere, if iron condensed into dust grains that were then removed from the atmosphere by a radiatively-driven wind.  If so, then one should compare the star's current abundance patterns with those of more iron-rich RGB stars.  The majority of stars in $\omega$ Cen have [Fe/H] $\sim -1.7$ \citep{Marino:2011B}; if ROA~5701 was once one of them, then its initial CNO abundance was $\epsilon$(C+N+O) = 7.69, rather than 7.43.  I assume the higher value in this analysis.

\subsection{The UV-Bright Stars in NGC 5904 (M5)}\label{sec_m5}

V42 was studied spectroscopically by \citet{Gonzalez:Lambert:1997} and \citet{Carney:1998}, who found that the star is not enhanced in $s$-process elements. It does show evidence for mixing on the RGB, particularly in its Na/O ratio; however, the star falls on the [Na/Fe] vs.\ [O/Fe] relation established by the cluster's RGB stars (Fig.\ 3 of \citeauthor{Carney:1998}), so additional mixing on the AGB is not necessary.  The star's lithium abundance is super-solar.  The production of lithium in AGB stars is generally explained via the Cameron-Fowler mechanism \citep{Cameron:Fowler:1971}, which is associated with the helium-shell flash.  This mechanism produces copious amounts of $s$-process elements; as these are not seen, some other process must be at work in V42.

V84 was studied by \citet[hereafter GL97]{Gonzalez:Lambert:1997}.  Again, the star exhibits no $s$-process enhancements.  GL97 found that the star's Na abundance is much greater than is seen in the cluster's RGB stars.  They suggested that the deep mixing responsible for the Na/O anticorrelation on the RGB continued on the AGB, further increasing the star's Na abundance. \citet{Carney:1998} pointed out that the Na line measured by GL97 is a blend.  For V42, they derived a value of [Na/Fe] that is 0.53 dex lower than that of GL97 for the same star.  If one subtracts 0.53 dex from the [Na/Fe] ratio quoted by GL97 for V84, then V84 falls near V42 on the cluster's [Na/Fe] vs.\ [O/Fe] relation.

\subsection{PAGB 2 in NGC 5986}\label{sec_pagb2}

\citet{Jasniewicz:2004} observed two stars in NGC~5986, PAGB 1 and PAGB 2 (ID 6 and ID 7, respectively).  They derived abundances for 18 elements in the spectrum of PAGB 2.  They found enhancements (relative to solar values) in the $s$-process elements that are similar to, but slightly lower than, those of ROA 24 in $\omega$ Cen.  The star is also enhanced in Ba and La.  These absorption features are weak or absent in the spectrum of PAGB 1.  Comparing the spectra of PAGB 1 and 2, the authors concluded that PAGB 2 experienced 3DU on the AGB.  (It was not possible to compare the star's abundances to those of RGB stars in NGC 5986, because no such abundance studies were available, a situation that has not improved.)

\subsection{K648 in NGC 7078 (M15)}\label{sec_k648}

K648 is the well-studied central star of the planetary nebula Ps 1 in M15 \citep{Alves:2000, Bianchi:2001, Rauch:2002}.  It is one of four PNe in Galactic globular clusters, and the only one for which detailed abundance measurements are available.  \citet{Otsuka:2015} derive the abundances of seven elements in the central star and ten in the nebula, and the gas and dust masses of the nebula.  Besides being enhanced in carbon, the star is more oxygen rich than the RGB stars in M15 by $\gtrsim 0.9$ dex.  \citet{Lugaro:2012} showed that stars with metallicities similar to M15's can synthesize significant quantities of both O and Ne during the thermal-pulsing AGB phase.  

 \citet{Otsuka:2015} concluded that the observed chemical abundances of the star and nebula and the gas mass of the nebula are in good agreement with the predictions of AGB nucleosynthesis models for stars with initial masses of 1.25 to 1.5 \msun.  Given its high core mass (0.61 to 0.63 \msun\ by their estimate), they suggest that K648 evolved from a progenitor that experienced coalescence or tidal disruption during the early stages of its evolution and became a 1.25--1.5 \msun\ blue straggler.
 
\subsection{Stars Omitted from this Sample}\label{sec_omitted}

One star not included in this sample is ZNG~1 in M5.  Based on observations with the {\em Far Ultraviolet Spectroscopic Explorer} (\fuse), \citet{Dixon:2004} derived a rotational velocity \vrot\ = $170 \pm 20$ \kms. They found the photosphere to be helium-rich and enhanced in carbon, nitrogen, and oxygen.  It is likely that the star represents the merger of two He-core white dwarfs (W.\ Dixon, in preparation).  If so, then its abundances cannot tell us about mixing and mass loss on the AGB.

Another excluded object is Y453 in NGC 6121.  Its surface gravity, \logg\ = 5.7, is sufficiently high that gravitational settling would quickly remove all metals from the photosphere; other processes, such as radiative levitation or a weak stellar wind, must be at work.  The star is thus a probe of these diffusion effects, rather than the thermal-pulsing AGB \citep{Dixon:2017}.

\subsection{Post-AGB stars as Standard Candles}\label{sec_candle}

\citet{Bond:1997A, Bond:1997B} suggested that optically-bright, intermediate-temperature post-AGB stars of old stellar populations can be used as extragalactic standard candles.  An upper limit to their luminosity distribution is set by the stars' evolutionary timescales: the most luminous post-AGB stars evolve across the Hertzsprung--Russell (H--R) diagram so quickly that they are bright at visible wavelengths for only a short time---even shorter if they leave the AGB while enshrouded in dust.  A lower limit to their luminosity distribution is set by the finite age of the universe: the lower-mass stars that will one day become fainter post-AGB stars have not yet left the main sequence.  

To explore this idea, \citet{Davis:2022} surveyed 97 Galactic globular clusters and identified 13 candidate post-AGB stars.  \citet{Ciardullo:2022} refined the sample, presenting ten post-AGB stars with $-0.05 \le (B - V)_0 \le 1.0 $ and $M_V$ brighter than $-3.0$.  (Six of my stars are among the \citeauthor{Ciardullo:2022} sample.)  The stars span a narrow range of luminosities: the scatter about a theoretical post-AGB evolutionary track with $\log (L/L_{\sun}) = 3.25$ is only 0.4 dex.  The stars are well separated in luminosity from other post-HB objects, particularly on the blue side of the instability strip.  Of the 438 post-HB stars in the \citeauthor{Davis:2022} catalog, no stars with colors  $0.0 \le (B - V)_0 \le 0.5 $ lie within $\sim$ 0.7 mag of the post-AGB track.

How do these results compare with those of the present study?  As illustrated in \figref{fig_mass}, 12 of my 17 stars have luminosities $\log (L/L_{\sun}) \sim 3.25$.  Of the five outliers, the two super-luminous stars have effective temperatures \teff\ $> 36,000$ K, so would not appear in surveys like that of \citet{Davis:2022}.  Two of the sub-luminous stars are the Bright Star in NGC 104 (47~Tuc) and ZNG~1 in NGC~6712.  Their colors, $(B - V)_0 = 0.14$ and $-0.12$, respectively \citep{Stetson:2019}, place them at the blue edge of the \citet{Ciardullo:2022} sample.    Using these stars to infer the distance to an external galaxy would bias the result.  The third sub-luminous star, V29 in $\omega$ Cen, is a magnitude fainter than the \citeauthor{Ciardullo:2022}\ sample.

\subsection{Horizontal-Branch Morphology}\label{sec_hbr}

\citet{Davis:2022} and \citet{Ciardullo:2022} pointed out that all of the post-AGB stars in their samples belong to clusters with blue HBs. They attributed this pattern to the strong dependence of post-AGB lifetime upon stellar mass.  Stars on the red HB have more massive envelopes, so their progeny have higher post-AGB masses and shorter post-AGB lifetimes.  To see if this result holds for my clusters, I employ the horizontal-branch ratio (HBR) of \citet{Lee:1994}.  Defined as HBR = (B$-$R) / (B+V+R), where B, V, and R are the numbers of BHB stars, RR Lyrae variables, and RHB stars, respectively, the ratio provides a measure of the color distribution of stars along the HB.  Positive values indicate clusters with mostly BHB stars, while negative values indicate clusters with mostly RHB stars. As shown in Table \ref{tab:clusters}, all but two of my clusters have BHBs.

\subsection{Metallicity Effects}\label{sec_metallicity}

\citet{Kamath:2023} computed post-AGB evolution models for stars with [Fe/H] = $-1.3$ and found that stars with ZAHB masses $\la 1$ \msun\ experience few thermal pulses, and their final surface composition shows little (if any) carbon and $s$-process enhancement.  \citet{Lugaro:2012} computed models with [Fe/H] = $-2.3$.  They found that stars with main-sequence masses as low as 0.9 \msun\ leave the AGB with surfaces enhanced in both C and O.  As pointed out by \citet{Karakas:2014}, two effects are at work.  First, the efficiency of 3DU increases with decreasing metallicity.  Second, because carbon is a product of the triple-$\alpha$ process, the amount of primary carbon generated during each thermal pulse does not depend on the metallicity of the star. In a low-metallicity star, the initial amount of CNO in the envelope is lower, so fewer pulses are required to enhance the CNO abundance.  All of the 3DU stars in this sample have [Fe/H] $< -1.65$ (Table \ref{tab:stars}).  On the other hand, only half of the stars with [Fe/H] $< -1.65$ show evidence of 3DU.

\subsection{Evolutionary Tracks}\label{sec_cmd}

\begin{figure}
\epsscale{1.1}
\plotone{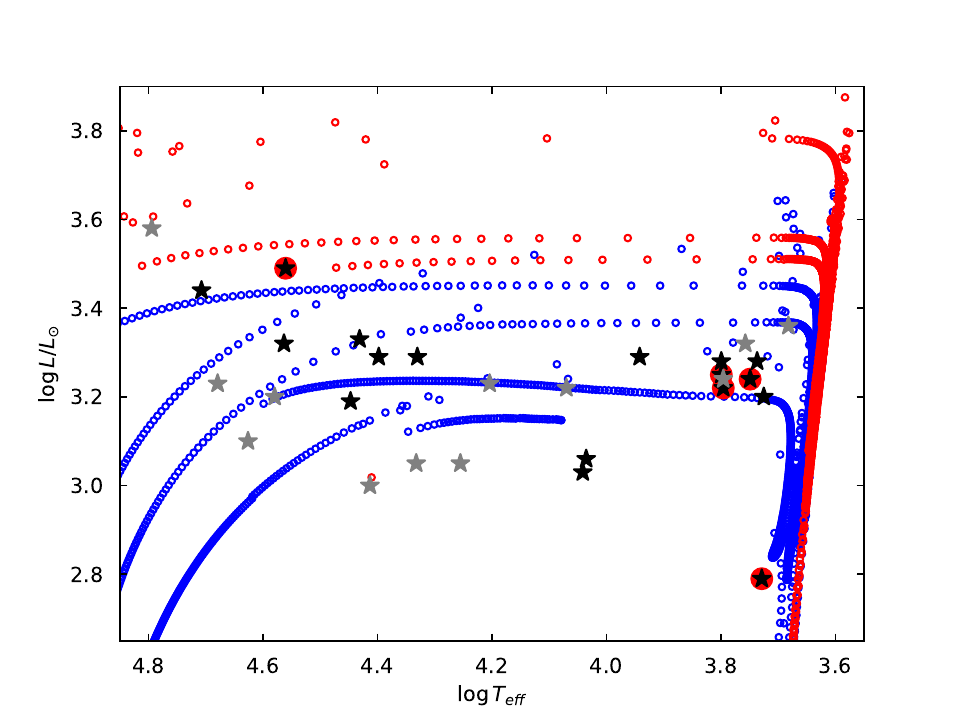}
\caption{Theoretical Hertzsprung--Russell diagram.  Stars from this sample are plotted in black; those exhibiting evidence of 3DU are highlighted in red.  Additional stars from the literature are plotted in grey.  Overplotted are post-HB evolutionary tracks for stars with [Fe/H] $= -1.5$ from \citet{Moehler:2019}.  Tracks for stars evolving from the BHB are represented by blue dots, those evolving from the RHB by red dots.  Each dot represents a time step of 1 Myr.}
\label{fig_cmd}
\end{figure}

In \figref{fig_cmd} the present sample of stars is plotted on the theoretical H--R diagram (black points).  To these are added the full samples of \citet{Moehler:2019} and \citet{Ciardullo:2022} (gray points).  Post-HB evolutionary tracks for stars with [Fe/H] $= -1.5$ from \citet{Moehler:2019} are overplotted.  Blue points represent the tracks for stars that evolved from the BHB (with ZAHB masses between 0.53 and 0.65 \msun).  Red points represent the tracks for more massive stars that evolved from the RHB.  Most post-AGB stars have temperatures and luminosities consistent with evolution from the BHB.

The evolutionary tracks depend sensitively on the stellar metallicity.  A plot similar to \figref{fig_cmd}, made using the [Fe/H] = $-2.0$ tracks from \citet{Moehler:2019}, shows that the most-massive BHB track (with $M_{\mathrm {ZAHB}} = 0.70$ \msun) passes between the two most luminous stars in my sample, ROB 162 in NGC 6397 and K648 in M15.  The least-massive RHB track is considerably brighter.  Low-metallicity models are more appropriate for these two clusters, which have [Fe/H] $< -2.0$.  Again, both stars' parameters are consistent with evolution from the BHB.

Among the grey points, the hot, luminous star in the upper left of the diagram is the central star of the planetary nebula GJJC-1 in NGC 6656 \citep[M22;][]{Gillett:1989}.  The star has not been the subject of an abundance analysis; it may well exhibit evidence of 3DU.  M22 is an intermediate-metallicity cluster ([Fe/H] $= -1.64$; \citealt{Harris:96, Harris:2010}) with a blue HB (HBR = 0.91; \citealt{Borkova:Marsakov:2000}), so the star's high luminosity suggests that it, too, may be a merger remnant \citep[cf.][]{Jacoby:2017}.  The ten points (both black and grey) with \teff\ $<$ 10,000 K and $\log (L/L_{\sun}) \ga 3.2$ are from the sample of \citet{Ciardullo:2022}.  The three new sub-luminous stars, ZNG 2 in NGC 6779, UIT 644 in NGC 5139, and IV-9 in NGC 6723 \citep{Moehler:2019}, are probably evolving along a low-mass BHB track.  Both NGC 6779 and NGC 5139 have BHBs, while NGC 6723 has stars on both sides of the instability strip (HBR = $-0.08$; \citeauthor{Borkova:Marsakov:2000}).

The least-luminous star in my sample, V29 in $\omega$ Cen, lies near the evolutionary track for a BHB star with $M_{\mathrm {ZAHB}} = 0.55$ \msun.  A star on this track will increase in luminosity to a value near $\log (L/L_{\sun}) = 3.2$.  I estimate a mass of 0.20 \msun\ for V29 (Table \ref{tab:stars}); if the actual value is closer to 0.55 \msun, then the star may eventually pass through the realm of the \citet{Ciardullo:2022} sample.

\subsection{Helium-Burning Stars}\label{sec_helium}

The clusters NGC 104 (47~Tuc) and NGC~6712 have RHBs, consistent with their high metallicities (Table \ref{tab:clusters}).  The progeny of RHB stars should be quite luminous, yet the Bright Star in 47~Tuc and ZNG~1 in NGC~6712 are among the faintest in my sample.  
\citet{Dixon:2021} suggested that the Bright Star is faint because it is burning helium instead of hydrogen.  In stellar-evolutionary calculations, whether a model departs the AGB as a H-burner or a He-burner depends on the timing of the last thermal pulse. Models departing the AGB during the quiescent H-burning phase (between thermal pulses) evolve as H-burners, while models departing during or immediately after a thermal pulse evolve as He-burners.  
 \citeauthor{Dixon:2021} found that the Bright Star is 0.3 dex fainter than is predicted for post-AGB H-burning stars.  For models with [Fe/H] = $-1.0$, the lowest-mass H-burning RHB track (with $M_{\mathrm {ZAHB}} = 0.65$ \msun) has a luminosity $\log (L/L_{\sun}) \sim 3.4$ \citep{Moehler:2019}, some 0.4 dex brighter than ZNG~1 in NGC~6712.  Thus, ZNG~1 may also be a He-burning object. 
 
Even during the thermal-pulsing phase, AGB stars spend far more time between thermal pulses than in them; nevertheless, the fact that this small sample may include two He-burning stars suggests that the probability of producing a sub-luminous, He-burning, post-AGB star is non-negligible.  Efforts to quantify this probability would facilitate the use of post-AGB stars as standard candles.

\section{Conclusions}\label{sec_conclusions}

\citet{Stasinska:2006} found that about 70\% of the post-AGB stars in their sample have experienced 3DU.  Roughly one-fourth of the stars in the present sample exhibit evidence of 3DU.  \citeauthor{Stasinska:2006}\ found a dependence of 3DU on metallicity, but not on mass.  Specifically, 3DU is more efficient at low metallicity, and the mass distribution of stars that did not experience 3DU is indistinguishable from those that did.  While all of the 3DU stars in this sample have [Fe/H] $< -1.65$, only half of the stars with [Fe/H] $< -1.65$ show evidence of 3DU.  The likelihood of 3DU seems to be independent of stellar mass.

I expected to find stellar luminosity correlated with stellar mass, and the degree of carbon enrichment correlated with both quantities, among post-AGB stars.  The majority of stars show neither trend.  This result may reflect both the large error bars on the stellar masses and the narrow range of progenitor masses among the sample stars.  The 3DU stars do exhibit a correlation between luminosity and mass, but the masses of V29 in $\omega$ Cen and K648 in M 15 may have been altered by binary interactions.  The relationship between carbon enrichment and luminosity among 3DU stars depends strongly on the properties of a single object, K648 in M15.  

The question now is, why do some post-AGB stars show evidence of 3DU, while others with similar mass and luminosity---and from clusters with similar metallicity and HB morphology---do not?  A star's post-HB evolutionary path is thought to be a sensitive function of its initial position on the HB (determined by its core and envelope masses), with metallicity and helium abundance playing a secondary role.  Are there trends in abundance or structure that predispose a star to experience 3DU, or is the process sufficiently stochastic as to be unpredictable?

From the observer's point of view, rather than looking for trends with luminosity, perhaps one should think in terms of three luminosity bins.  Most globular-cluster post-AGB stars fall in the range $3.15 \lesssim \log L/L_{\sun} \lesssim 3.35$.  They are the progeny of blue HB stars in clusters with intermediate metallicity ([Fe/H] $\sim -1.5$).  A second group consists of sub-luminous stars in high-metallicity clusters ([Fe/H] $\sim -1.0$) with red HBs.  They may be burning helium, rather than hydrogen.  A third group of hot, super-luminous stars is evolving quickly across the Hertzsprung--Russell diagram.  Some of them may be merger remnants.

\begin{acknowledgments}

The author is indebted to M.~M.\ Miller Bertolami for thoughtful comments on the manuscript.
This work has made use of NASA's Astrophysics Data System (ADS) and the SIMBAD database, operated at CDS, Strasbourg, France.  It was performed during research leave granted by the STScI Science Mission Office, whose support is gratefully acknowledged.  Its publication is supported by the STScI Director's Discretionary Research Fund.  

\end{acknowledgments}

%

\vspace{5mm}


\software{MASH3 \citep{Worthey:2011}}





\end{document}